\documentclass[11pt,a4paper,english,amssymb,nofootinbib,superscriptaddress]{revtex4}
\usepackage{lmodern}

\usepackage[T1]{fontenc}
\usepackage[latin9]{inputenc}
\usepackage{babel}
\usepackage{amsmath}
\usepackage{graphicx}
\usepackage{amssymb}
\usepackage{esint}
\usepackage{color}
\usepackage{soul}

\usepackage[unicode=true, pdfusetitle,
bookmarks=true,bookmarksnumbered=false,bookmarksopen=false,
breaklinks=false,pdfborder={0 0 1},backref=false,colorlinks=false]
{hyperref}
\def\b{\begin{equation}}
\def\e{\end{equation}}

\makeatletter
\@ifundefined{textcolor}{}
{%
 \definecolor{BLACK}{gray}{0}
 \definecolor{WHITE}{gray}{1}
 \definecolor{RED}{rgb}{1,0,0}
 \definecolor{GREEN}{rgb}{0,1,0}
 \definecolor{BLUE}{rgb}{0,0,1}
 \definecolor{CYAN}{cmyk}{1,0,0,0}
 \definecolor{MAGENTA}{cmyk}{0,1,0,0}
 \definecolor{YELLOW}{cmyk}{0,0,1,0}
 }

\usepackage{latexsym}\usepackage{bm}

\makeatother

\makeatother

\begin{document}
\title{$pp$-waves as exact solutions to ghost-free infinite derivative gravity}
\author{Ercan Kilicarslan}

\email{ercan.kilicarslan@usak.edu.tr}

\affiliation{Department of Physics,\\
 Usak University, 64200, Usak, Turkey}

\begin{abstract}
We construct exact pp-wave solutions of ghost-free infinite derivative gravity and demonstrate that the sourceless theory does not bring any pp-wave solutions save for that of Einstein's gravity. These waves described in the  Kerr-Schild form also solve the linearized field equations of the theory. We also find an exact gravitational shock wave with non-singular curvature invariants and with a finite limit in the ultraviolet regime of non-locality which is in contrast to the divergent limit in Einstein's theory. 
\end{abstract}

\maketitle

\section{Introduction}
Among the small scale modifications of Einstein's theory of General Relativity (GR), infinite derivative gravity (IDG) \cite{Biswas2, Biswas1,Modesto1} seems to be a viable candidate to have a complete theory in the UV scale (short distances). A particular form of IDG is free from the Ostragradsky type instabilities and black hole or cosmological type singularities. The theory is described by a Lagrangian density built from analytic form factors which lead to non-local interactions. The propagator of ghost and singularity free  IDG in flat background has obtained by modification of pure GR propagator via an exponential of an entire function which has no roots in the finite domain \cite{Biswas1,Biswas:2013kla}. This modification provides that the theory does not have ghost-like instabilities and extra degree of freedom (DOF) other than the massless graviton. On the other hand, infinite derivative extension of GR describes non-singular Newtonian potential for a point-like source at small distances \cite{Biswas1,Edholm:2016hbt}. This result is extended to the case where point-like sources also have velocities, spins and orbital motion which leads to spin-spin and spin-orbit interactions in addition to mass-mass interactions \cite{Kilicarslan:2018yxd}. It was shown that not only mass-mass interaction but also spin-spin and spin orbit interactions are non-singular in the UV regime of non-locality. Hence, the theory is well-behaved in the small scale unlike GR. Furthermore, power counting arguments have been recently studied for renormalizability discussion and it is shown that loop-diagrams beyond one-loop may give finite result with dressed propagators \cite{Modesto1,Talaganis,Talaganis:2016ovm,Tomboulis,Buoninfante:2018mre,Calcagni:2014vxa}. Moreover, IDG maybe devoid of black hole and cosmological Big Bang type singularities at a linear and non-linear level \cite{Biswas1, Biswas2,Tomboulis,Biswas4, Modesto,Biswas5,Biswas6, Frolov:2015bta, Koshelev:2017tvv,Koshelev:2018hpt,Boos:2018bxf,Boos:2018bhd}.  These encouraging developments led us to study exact solutions of the theory.

There are many works and some books on finding and classifying the exact solutions of Einstein's gravity \cite{book}. Furthermore, some exact solutions are studied in detail in some specific modified gravity theories, such as the quadratic gravity \cite{Eddington,Buchdahl,Buchdahl2,Gullu:2011sj,Gurses:2012db,Gurses:2015zia}, higher order theories of gravity \cite{Gursespp}, $f(Riemann)$ theories \cite{Gurses:2013jua}, $f(R_{\mu\nu})$ theories \cite{Gurses:2011fv} and $f(R)$ theories \cite{Sotiriou:2008rp}. On the other hand, although IDG received attention in the recent literature, exact solutions of the theory have not been studied at a non-linear level\footnote{Some exact solutions of weakly non-local gravity theories are discussed in \cite{Li:2015bqa}.} since the field equations are very lengthy and complicated. At the linearized level around a flat background, some specific solutions have been found: a non-singular rotating solution without ring singularity was studied in \cite{Mazumdar1}, a solution for an electric point charge was found in \cite{Mazumdar2}, conformally flat static metric was constructed in \cite{Mazumdar3}, a metric for the non-local star was given in \cite{Buoninfante:2019swn}. However, at the non-linear level, we are not aware of any known exact solution for the theory. Nevertheless, since Kundt Einstein spacetimes of Petrov 
(Weyl) type N are universal \cite{Gueven:1987ad,Amati,Horowitz, Coley, Hervik}, these spacetimes are exact solutions of IDG. 

 In this work, we would like to construct exact pp-wave solutions of the IDG. Therefore, we consider the pp-wave metric in the Kerr-Schild form which leads to remarkable simplification in finding exact solutions. We show that pp-wave spacetimes are exact solutions of the IDG. We also show that these waves solve not only generic non-linear field equations but also the linearized ones. Furthermore, pp-wave solutions of Einstein's theory also solve the IDG since they are Kundt spacetimes of Petrov type-N with zero curvature scalar \cite{Gueven:1987ad,Amati,Horowitz, Coley, Hervik}. We also discuss the pp-wave solution of the theory in the presence of the null matter which contains Dirac delta type singularity, namely we construct an exact non-singular gravitational shock-wave solution at the non-linear level. We show that curvature tensors are regular at the origin. Although, exact gravitational shock wave solution of Einstein's theory generated by massless point particle is singular at the origin, gravitational non-local interactions in IDG leads to cancellation of such a singularity at the non-linear level.

The layout of the paper is as follows: In Sec. II, we will briefly review the IDG. Sec. III is devoted to some mathematical preliminaries of the pp-wave metrics in the Kerr-Schild form.  In that section, we write the generic field equations of IDG for pp-wave spacetimes.  In Sec. IV, we give the explicit form of the exact solution for ghost-free IDG. In addition to the non-linear theory, we show that pp-wave solutions of the generic theory also satisfy the linearized field equations. In Sec. V, we construct the exact non-singular gravitational shock-wave solutions of IDG. 

\section{Infinite Derivative Gravity}
The most general quadratic, parity-invariant and torsion-free Lagrangian density of IDG is \cite{Biswas1, Biswas2, Modesto1}
\begin{equation}
\mathcal{L}= \frac{1}{16\pi G}\sqrt{-g}\bigg[\  R\ +\alpha_c\bigg( R {\cal
F}_1 (\Box) R +  R_{\mu\nu} {\cal
F}_2(\Box) R^{\mu\nu}
       + C_{\mu\nu\rho\sigma} {\cal
F}_3(\Box) C^{\mu\nu\rho\sigma}\bigg)\bigg],
\end{equation}
where $G=\frac{1}{M_p^2}$ is the Newton's gravitational constant and $\alpha_c=\frac{1}{M_s^2}$ is a dimensionful parameter where $M_s$ is the scale of the non-locality, $R$ is the scalar curvature, $R_{\mu\nu}$ is the Ricci tensor and $C_{\mu\nu\rho\sigma}$ is the Weyl tensor. We work with the $(-,+,+,+)$ signature. In the $\alpha_c \to 0$ (or $M_s \to \infty$) limit, the theory reduces to Einstein's gravity with a massless spin-$2$ graviton. Note that IDG is a special case of ghost-free quadratic curvature theories of gravity. On the other hand, the three form factors ${\cal
F}_i(\Box)$'s containing infinite derivative functions are defined as \footnote{These three form factors are not independent and are constrained. For example, in flat background to conserve general covariance and the massless spin-$2$ nature of graviton, these form factors satisfies the following constraint equation \cite{Biswas1,Biswas4}
\begin{equation}
6{\cal
F}_1(\Box)+3{\cal
F}_2(\Box)+2{\cal
F}_3(\Box)=0,
\label{constraint}
\end{equation}
which provides that theory has only transverse-traceless massless spin-$2$ graviton degree of freedom.} 
\begin{equation}
{\cal
F}_i(\Box)\equiv\sum_{n=0}^{\infty}f_{i_n}\frac{\Box^n}{M_s^{2n}}, 
\label{idfunc}
\end{equation} 
in which $f_{i_n}$ are dimensionless coefficients. The form factors lead to non-local gravitational interactions and $f_{i_n}$ play an important role to avoid ghost-like instabilities. The source-free field equations are \cite{Biswas4}
\begin{equation}
\begin{aligned}
&G^{\alpha\beta}+\frac{\alpha_c}{2}\bigg(4G^{\alpha\beta}{\cal
F}_{1}(\Box)R+g^{\alpha\beta}R{\cal
F}_1(\Box)R-4\left(\triangledown^{\alpha}\nabla^{\beta}-g^{\alpha\beta}
\square\right){\cal F}_{1}(\Box)R
\\&
+4R^{\alpha}\,_{\nu}{\cal F}_2(\Box)R^{\nu\beta}
-g^{\alpha\beta}R_{\nu}\,^{\mu}{\cal
F}_{2}(\Box)R_{\mu}\,^{\nu}-4\triangledown_{\nu}\triangledown^{\beta}({\cal
F}_{2}(\Box)R^{\nu\alpha})
+2\square({\cal
F}_{2}(\Box)R^{\alpha\beta})
\\&
+2g^{\alpha\beta}\triangledown_{\mu}\triangledown_{
\nu}({\cal F}_{2}(\Box)R^{\mu\nu})
-g^{\alpha\beta}C^{\mu\nu\rho\sigma}{\cal
F}_{3}(\Box)C_{\mu\nu\rho\sigma}+4C_{\;\mu\nu\sigma}^{\alpha}{\cal {\cal
F}}_{3}(\square)C^{\beta\mu\nu\sigma}
\\&
-4(R_{\mu\nu}+2\triangledown_{\mu}
\triangledown_{\nu})({\cal {\cal F}}_{3}(\square)C^{\beta\mu\nu\alpha})
-2\Omega_{1}^{\alpha\beta}+g^{\alpha\beta}(\Omega_{1\rho}^{\;\rho}+\bar{
\Omega}_{1}) 
-2\Omega_{2}^{\alpha\beta}+g^{\alpha\beta}(\Omega_{2\rho}^{\;\rho}+\bar{
\Omega}_{2})\\& -4\Delta_{2}^{\alpha\beta}
-2\Omega_{3}^{\alpha\beta}+g^{\alpha\beta}(\Omega_{3\gamma}^{\;\gamma}+\bar{
\Omega}_{3}) -8\Delta_{3}^{\alpha\beta}\bigg)=0.
\label{IDGfeqns}
\end{aligned}
\end{equation}
Here, the symmetric tensors are given as \cite{Biswas4}
\begin{equation}
\begin{aligned}
&\Omega_{1}^{\alpha\beta}=\sum_{n=1}^{\infty}f_{1_{n}}\sum_{l=0}^{n-1}\nabla^{
\alpha}R^{(l)}\nabla^{\beta}R^{(n-l-1)},\quad\bar{\Omega}_{1}=\sum_{n=1}^{\infty
}f_{1_{n}}\sum_{l=0}^{n-1}R^{(l)}R^{(n-l)},\\&
\Omega_{2}^{\alpha\beta}=\sum_{n=1}^{\infty}f_{2_{n}}\sum_{l=0}^{n-1}R_{\nu}\,^{
\mu;\alpha(l)}R_{\mu}\,^{\nu;\beta(n-l-1)},\quad\bar{\Omega}_{2}=\sum_{n=1}^{
\infty}f_{2_{n}}\sum_{l=0}^{n-1}R_{\nu}\,^{\mu(l)}R_{\mu}\,^{\nu(n-l)}\\&
\Delta_{2}^{\alpha\beta}=\frac{1}{2}\sum_{n=1}^{\infty}f_{2_{n}}\sum_{l=0}^{n-1}
[R_{
\;\sigma}\,^{\nu(l)}R^{(\beta|\sigma|;\alpha)(n-l-1)}-R_{\;\sigma}\,^{\nu;(\alpha(l)
}R^{
\beta)\sigma(n-l-1)}]_{;\nu}\\&
\Omega_{3}^{\alpha\beta}=\sum_{n=1}^{\infty}f_{3_{n}}\sum_{l=0}^{n-1}C_{
\;\nu\rho\sigma}^{\mu;\alpha(l)}C_{\mu}\,^{\;\nu\rho\sigma;\beta(n-l-1)},
\;\bar{\Omega}_{3}=\sum_{n=1}^{\infty}f_{3_{n}}\sum_{l=0}^{n-1}C_{
\;\nu\rho\sigma}^{\mu(l)}C_{\mu}^{\;\nu\rho\sigma(n-l)}\\&
\Delta_{3}^{\alpha\beta}=\frac{1}{2}\sum_{n=1}^{\infty}f_{3_{n}}\sum_{l=0}^{n-1}
[C_{
\;\;\;\sigma\mu}^{\rho\nu(l)}C_{\rho}^{\;(\beta|\sigma\mu|;\alpha)(n-l-1)}
-C_{
\;\;\;\sigma\mu}^{\rho\nu\;\;;(\alpha(l)}C_{\rho}^{
\;\beta)\sigma\mu(n-l-1)}
]_{;\nu}
\end{aligned}
\end{equation}
where we used the notation $R^{(n)}=\Box^n R$ for the tensors which are built from the curvature tensors and their derivatives and semi-colon denotes covariant derivative. Note that since the field equations are highly complicated and non-linear, finding exact solutions to the theory might seem hopeless. In the next section, we will give some mathematical preliminaries of the pp-wave spacetimes and show that these spacetimes are exact solution of the theory for a proper choice of the profile function.
\section{PP-wave spacetimes in IDG}
Here we want to find the pp-wave solution of the theory.  For this purpose, let us consider the pp-wave (or plane-fronted parallel waves) metric described in the Kerr-Schild form as \footnote{For the detailed properties of pp-waves, see \cite{Blau,Classification,Gursespp,Edelstein:2016nml}. }
\begin{equation}
g_{\mu \nu}=\eta_{\mu \nu}+2 H \lambda_\mu \lambda_\nu.
\label{KSForm}
\end{equation}
 Here $\eta_{\mu \nu}$ denotes the flat metric and the covariantly constant null vector $\lambda_\mu$ satisfies the following relations
\begin{equation}
\begin{aligned}
\lambda^{\mu}\lambda_\mu=0,\hskip .6 cm\nabla_{\mu}\lambda_\nu=0,\hskip .6 cm
\label{plane1}
\end{aligned}
\end{equation}
which give $\lambda^\mu\partial_\mu H=0$. The null vector $\lambda_\mu$ is non-expanding $\nabla_{\mu}\lambda^\mu=0$, non-twisting $\nabla_{\mu}\lambda^\nu\nabla_{[\mu}\lambda_{\nu]}=0$  and shear-free $\nabla_{\mu}\lambda^\nu\nabla_{(\mu}\lambda_{\nu)}=0$, hence the pp-wave metrics belong to class of the Kundt spacetimes \cite{book}. The inverse metric reads as
\begin{equation}
g^{\mu \nu}=\eta^{\mu \nu}-2 H \lambda^\mu \lambda^\nu.
\end{equation}
To find the pp-wave solution of IDG, one needs to calculate relevant tensors (such as the Riemann, Ricci and  scalar curvature) corresponding to metric. For this purpose, let us note that the Christoffel connection can be computed to be 
\begin{equation}
\Gamma^\sigma_{\mu \nu}= \lambda^\sigma \lambda_\mu \partial_\nu H + \lambda^\sigma \lambda_\nu \partial_\mu H - \lambda_\mu \lambda_\nu \eta^{\sigma \beta} \partial_\beta H, 
\label{ConnectionPP}
\end{equation}
which satisfies $\lambda_\sigma\Gamma^\sigma_{\mu \nu}=0$, $\lambda^\mu\Gamma^\sigma_{\mu \nu}=0$. Now we are ready to calculate Riemann, Ricci and Weyl tensors. The Riemann tensor can be found as \cite{Horowitz}
\begin{equation}
R_{\rho\sigma\mu\nu}=\lambda_{\rho}\lambda_{\nu}\partial_{\sigma}\partial_{\mu}H+\lambda_{\sigma}\lambda_{\mu}\partial_{\rho}\partial_{\nu}H
-\lambda_{\rho}\lambda_{\mu}\partial_{\sigma}\partial_{\nu}H-\lambda_{\sigma}\lambda_{\nu}\partial_{\rho}\partial_{\mu}H,
\label{Riemannpp}
\end{equation}
with which one gets the Ricci tensor as
\begin{equation}
R_{\mu\nu}=-\lambda_{\mu}\lambda_{\nu}\partial^2 H,
\label{RicciPP}
\end{equation}
where $\partial^2$ is flat space Laplace operator defined as $\partial^2=\eta^{\mu\nu}\partial_\mu\partial_\nu$. It is straightforward to see that scalar curvature is zero as a consequence of the fact that  $\lambda_\mu$ is null. Note that any contraction of $\lambda_\mu$ with Weyl, Riemann and Ricci tensors vanishes: 
\begin{equation}
\lambda^\mu C_{\rho\sigma\mu\nu}=0, \hskip .5 cm  \lambda^\mu R_{\rho\sigma\mu\nu}=0, \hskip .5 cm \lambda^\mu R_{\mu\nu}=0.
\end{equation}
Furthermore, all the curvature scalars vanish for the pp-wave metric \cite{Peres,Pravda:2002us}.
On the other hand, the pp-waves have some remarkable algebraic properties which provide simplicity in calculations. For example, any non-trivial second rank tensor built from Riemann tensor or its covariant derivatives can be described by a linear combination of traceless-Ricci \footnote{By traceless Ricci tensor, we mean $S_{\mu\nu}\equiv R_{\mu\nu}-\frac{1}{4}g_{\mu\nu}R$ where $S_{\mu\nu}$ is traceless Ricci tensor.}  and higher-orders of traceless-Ricci ($\Box^n S_{\mu\nu}$'s) tensors \cite{Gursespp}. With this property and vanishing of all scalar invariants, the pp-wave spacetimes are Weyl type N. Another remarkable property of the pp-wave metric is that contraction $\lambda_\mu$ vector with $\nabla^{n}H$'s  vanish \cite{Gursespp}   
\begin{equation}
\begin{aligned}
\lambda^{\mu_{1}}\nabla_{\mu_{1}}\nabla_{\mu_{2}}\dots\nabla_{\mu_{n}}H =0,
\end{aligned}
\end{equation}
which will be frequently used in the paper. Therefore, $\lambda$ contraction with other $\lambda$'s or with $\nabla^{n}H$'s give zero.  Finally, let us consider the structure of a non-zero term given in the form
\begin{equation}
\begin{aligned}
\nabla_{\nu_{1}}\nabla_{\nu_{2}}\dots\nabla_{\alpha}\dots\nabla_{\beta}\dots\nabla_{\nu_{2n-2}}C^{\beta\mu\alpha\nu}=\frac{1}{2}\nabla_{\nu_{1}}\nabla_{\nu_{2}}\dots\nabla_{\nu_{2n-2}}\square R^{\mu\nu},
\end{aligned}
\end{equation}
where we have used the following twice-contracted Bianchi identity of the Weyl tensor for the pp-wave metric (\ref{KSForm})
\begin{equation}
\nabla_{\alpha}\nabla_{\beta}C^{\beta\mu\alpha\nu}=\frac{1}{2}\square R^{\mu\nu}.
\end{equation}
\subsection{Field equations of the IDG for pp-wave spacetime}
Now we are ready to write the field equations of the IDG for the pp-wave spacetimes. By using relations obtained above for each term in the field equations, thanks to the fact that pp-waves have a Riemann tensor of type N together with all its derivatives (and also $R=0$), only terms linear in the curvature give non-zero contribution in (\ref{IDGfeqns}) \cite{Horowitz, Hervik}, the field equations take the form
\begin{equation}
\bigg[1+\alpha_c \bigg(\Box {\cal F}_{2}(\Box)+2 {\cal F}_{3}(\Box)\Box\bigg)\bigg]R_{\mu\nu}=0.
\label{ppgeneq}
\end{equation}
Note that the pp-wave metrics which satisfy $R_{\mu\nu}=0$ also solve IDG field equations (\ref{ppgeneq}).
Using Ricci tensor definition (\ref{RicciPP}) for pp-wave metric ansatz, the complete field equations (\ref{ppgeneq}) can be recast as
\begin{equation}
\bigg[1+\alpha_c \bigg(\Box {\cal F}_{2}(\Box)+2{\cal F}_{3}(\Box)\Box \bigg)\bigg]\partial^2H=0,
\label{difeq}
\end{equation}
where we also used the fact that the null vector is covariantly constant. Since the form factor ${\cal F}_{2}$ and ${\cal F}_{3}$ can be described in terms of generic operator of d'Alembert as
\begin{equation}
{\cal
F}_2(\Box)=\sum_{n=0}^{\infty}f_{2_n}\frac{\Box^n}{M_s^{2n}},\hskip .5 cm {\cal
F}_3(\Box)=\sum_{n=0}^{\infty}f_{3_n}\frac{\Box^n}{M_s^{2n}},
\end{equation} 
one needs to evaluate the $\square^{n}H$. For this purpose, first let us consider the box operator acting on $H$
\begin{equation}
\square H=g^{\mu\nu}\nabla_{\mu}\nabla_{\nu}H=\eta^{\mu\nu}\partial_{\mu}\partial_{\nu}H-\eta^{\mu\nu}\Gamma_{\mu\nu}^{\sigma}\partial_{\sigma}H.
\label{boxpp}
\end{equation}
By using Eq.(\ref{ConnectionPP}), it can be easily shown that the last term vanishes since $\eta^{\mu\nu}\Gamma_{\mu\nu}^{\sigma}=0$. Then the equation (\ref{boxpp}) takes the form 
\begin{equation}
\square H=\partial^{2}H.
\label{boxh}
\end{equation}
Consequently, one can show that $\Box^n\partial^{2}H=\partial^{2n}(\partial^{2}H)$, with which the field equations of IDG (\ref{difeq}) reduce to
\begin{equation}
\bigg[1+\alpha_c\bigg( \partial^{2} {\cal F}_{2}(\partial^{2})+2\partial^{2} {\cal F}_{3}(\partial^{2})\bigg)\bigg]\partial^{2}H=0,
\label{difeq1}
\end{equation}
whose most general solution can be given as
\begin{equation}
H_{IDG}=H_{E}+\Re\left(H_{I}\right),
\end{equation}
where $H_{E}$ refers to the solution of pure Einstein's gravity and satisfies the equation $\partial^{2}H_{E}=0$, $H_{I}$ is the solution to IDG theory
solving equation $\bigg[1+\alpha_c\bigg( \partial^{2} {\cal F}_{2}(\partial^{2})+2\partial^{2} {\cal F}_{3}(\partial^{2})\bigg)\bigg]H_{I}=0$ and $\Re$ denotes the real part of the solution of $H_{I}$. Here, one should notice that
the pp-wave metric solution of Einstein's theory also solves IDG theory. 

For the choice of the form factor ${\cal F}_{2}={\cal F}_{3}=0$ which yields the theory
\begin{equation}
\mathcal{L}= \frac{1}{16\pi G}\sqrt{-g}\bigg[\  R\ +\alpha_c( R {\cal
F}_1 (\Box) R 
       \bigg],
\end{equation}
which has non-singular bouncing solution which may avoid cosmological singularity problem \cite{Biswas2}. The associated field equations for the pp-wave spacetimes reduce to $\partial^{2}H=0$. This shows that the pp-wave solutions of Einstein theory are exact solution of the theory.
\section{pp-wave Solutions}
In order to obtain the explicit form of solution (\ref{difeq1}), one can describe the pp-wave metric in null coordinates with the appropriate choice of $\lambda_{\mu}$ as \cite{book}
\begin{equation}
ds^{2}=2dudv+2H\left(u,x,y\right)du^{2}+dx^2+dy^2,
\label{ppcoor}
\end{equation}
in which $u$ and $v$ are light-cone background coordinates defined as $u=\frac{1}{\sqrt{2}}(x-t)$ and $v=\frac{1}{\sqrt{2}}(x+t)$. Here, since $\lambda_{\mu}=\delta_\mu^{u}$ which yields $\lambda^{\mu}=\delta^\mu_{v}$, we have  
\begin{equation}
\lambda_{\mu}dx^{\mu}=du, \hskip .5 cm\lambda^{\mu}\partial_{\mu}H=\partial_{v}H=0.
\end{equation}
With these properties and using the Laplacian for the metric (\ref{ppcoor}) as $\partial^{2}=2\frac{\partial^{2}}{\partial u\partial v}+\partial_\perp^{2}$, here $\partial_\perp^{2}=\partial_x^{2}+\partial_y^{2}$, equation (\ref{boxh}) takes the form
\begin{equation}
\square H=\partial_\perp^{2}H,
\end{equation}
where we used the fact that  $\partial_{v}H=0$, and similarly one has,
\begin{equation}
\square^{n}H=\partial_\perp^{2n}H.
\end{equation}
and (\ref{difeq1}) reduces to 
\begin{equation}
\bigg[1+\alpha_c \bigg(\partial_\perp^{2}{\cal F}_{2}(\partial_\perp^{2})+2\partial_\perp^{2}{\cal F}_{3}(\partial_\perp^{2})\bigg)\bigg]\partial_\perp^{2}H=0,
\label{difeqppfinal}
\end{equation}
which is the general equation that we want to solve. To proceed further we need the explicit form of  form factors ${\cal F}_{2}(\Box)$ and ${\cal F}_{3}(\Box)$.
\subsection{Explicit Solutions}
For the sake of simplicity, one can choose the following form factors that satisfy ghost-freedom \cite{Biswas2,Biswas1}
\begin{equation}
 {\cal F}_{2}(\Box)=-2 {\cal F}_{1}(\Box)=\frac{-1+e^{-\frac{\Box}{M_s^2}}}{\frac{\Box}{M_s^2}},\hskip .5 cm {\cal F}_{3}(\Box)=0,
 \label{formfactor}
\end{equation}
which satisfies the constraint equation (\ref{constraint}).With this setting, the theory has only massless spin-2 graviton about the flat background. The corresponding field equation (\ref{difeqppfinal}) takes the form
\begin{equation}
e^{-\frac{\partial_\perp^{2}}{M_s^2}} \partial_\perp^{2}H=0.
\label{IDG}
\end{equation}
To solve this differential equation, even if one can also use the \textit{eigenvalue method} defined in \cite{BarnabyKamran}, here as demonstrated in \cite{BarnabyKamran,Kilicarslan2}, the solution of original equation is just given with the following equation
\begin{equation}\label{greq}
\partial_\perp^{2}H=0,
\end{equation}
which is exactly the field equation satisfied by the pp-wave solutions of Einstein's gravity. In other words, $pp$ wave solutions of the source-free Einstein's gravity is also the solutions of IDG. Notice that all the analytic solutions of (\ref{greq}) are very well-known \cite{book}. As an example, the gravitational plane wave solution of Einstein's theory is given as follows 
\begin{equation}
H(u,x,y)=A(u)(x^2-y^2)+B(u)xy,
\end{equation}
where $A(u)$ and $B(u)$ are any arbitrary smooth functions of null coordinate $u$. 
Observe that, as expected, the non-local interactions do not play any role in the source-free theory \cite{Kilicarslan2} since the field equations are linear for the $pp$-wave metric ansatz. To see the non-local effects, we will consider the null source coupled field equations in the Sec.\ref{secv}.

\subsection{Linearized Field equations of IDG as exact field equations}
In this part, we wish to consider the pp-wave solutions of the linearized form of IDG. In fact, one can recognize from (\ref{ppgeneq}) that the pp-wave metric solves both the full IDG field equations and the linearized version. In other words, by defining the metric perturbation $h_{\mu\nu}=g_{\mu\nu}-\eta_{\mu\nu}=2H\lambda_{\mu}\lambda_{\nu}$, the exact field equations of the IDG takes the form of the linearized field equations. To show this explicitly, let us turn our attention to the source-free linearized field equations of the IDG around the flat background of $g_{\mu\nu}=\eta_{\mu\nu}+h_{\mu\nu}$
\begin{equation}
a(\Box)R^L_{\mu\nu}-\frac{1}{2}\eta_{\mu\nu}c(\Box)R^L-\frac{1}{2}f(\Box)\partial_\mu\partial_\nu R^L=0,
\label{FeqIDG}
\end{equation}
where $L$ denotes the linearization and infinite derivative non-linear functions are described as
\begin{equation}
\begin{aligned}
 &a(\Box) =1 + M^{-2}_s \left({\cal{F}}_2(\Box) 
        + 2  {\cal{F}}_3(\Box)\right) \Box, \\
&        c(\Box) = 1 - M^{-2}_s\left(4 {\cal{F}}_1(\Box) +  {\cal{F}}_2(\Box)  
        - \frac{2}{3}{\cal{F}}_3(\Box)\right)\Box,\\
&        f(\Box) =M^{-2}_s \left(4{\cal{F}}_1(\Box) + 2{\cal{F}}_2(\Box) +\frac{4}{3} {\cal{F}}_3(\Box)\right),
        \end{aligned}
        \label{relations}
\end{equation}
which yield the constraint $a(\Box)-c(\Box) = f(\Box)\Box$. In the metric perturbation $h_{\mu\nu}=g_{\mu\nu}-\eta_{\mu\nu}=2H\lambda_{\mu}\lambda_{\nu}$ for the Kerr-Schild form, after using the linearized form of curvature tensors \cite{deser_tekin}, the linearized Ricci and scalar curvature will read, respectively
\begin{equation}
R^L_{\mu\nu}=-\frac{1}{2}\partial^2 h_{\mu\nu}=-H \lambda_{\mu}\lambda_{\nu}, \hskip .5 cm R^L=0.
\label{lineartensors}
\end{equation} 
Observe that the metric perturbation $h_{\mu\nu}$ is transverse-traceless: $h=0$ and $\nabla^{\mu}h_{\mu\nu}=0$, hence the linearized scalar curvature $R^L$ vanishes. Furthermore, the theory describes only massless transverse-traceless spin-$2$ DOF. Accordingly, by plugging the linearized tensors (\ref{lineartensors}) into the linearized field equations, one gets
\begin{equation}
a({\Box}) (\Box H) =0.
\label{Lin}
\end{equation}
To further reduce (\ref{Lin}), using the definition of non-linear function $a({\Box})$ (\ref{relations}), one gets
\begin{equation}
\bigg[1 + \alpha_c \bigg({\cal{F}}_2(\Box) 
        + 2  {\cal{F}}_3(\Box)\bigg) \Box\bigg](\Box H) =0.
\end{equation}
This shows that all solutions of the linearized field equations for the metric perturbation $h_{\mu\nu}$ satisfy the non-linear field equations of the IDG. Furthermore, the field equations of linearized theory coincide with non-linear theory for the pp-wave metric. Moreover, in order to have ghost freedom, $a({\Box})$ should be an entire function. The simplest choice is $a({\Box})=e^{-\frac{\Box}{M_s^2}}$\cite{Biswas1}.Thus, the field equations reduce to 
\begin{equation}
e^{-\frac{\Box}{M_s^2}} (\Box H) =0.
\end{equation}
For the metric (\ref{ppcoor}), the final result for the linearized field equations is
\begin{equation}
e^{-\frac{\partial_\perp^{2}}{M_s^2}} \partial_\perp^{2}H=0.
\end{equation}

\section{ Exact non-singular gravitational shock wave solution of IDG \label{secv}}
In this section, we would like to extend the pp-wave solutions in the presence of the pure radiation sources (null dust). Gravitational shock wave solution can provide understanding of the gravitational interactions between high energy massless particles in IDG. Shock waves are special class of axisymmetric pp-waves and its metric produced by a fast moving massless point particle can be described as follows \cite{book3,book2} \footnote{In fact we can use pp-wave metric given in the form (\ref{ppcoor}), but the form of equation (\ref{ansatz}) is commonly used in the literature. Therefore, we use this form. Note that the metric (\ref{ansatz}) can also be described in Kerr-Schild form as $g_{\mu \nu}=\eta_{\mu \nu}+ V\lambda_\mu \lambda_\nu$ which leads to $R_{\mu\nu}=-\frac{1}{2}\lambda_{\mu}\lambda_{\nu}\partial^2 V$ where $V=\delta(u)g(x_\perp)$. }
\begin{equation}
ds^2=-du dv+\delta(u)g(x_\perp) du^2+dx_\perp^2,\label{ansatz}
\end{equation}
where $u=t-z$ and $v=t+z$ are the null-cone background coordinates\footnote{$(t,x_\perp,z)$ be the coordinates in the Minkowski space.}, $(x^i)=x_\perp$ with $ i=1,2 $  are the transverse coordinates to wave propagation and $g(x_\perp)$ is the wave profile function. To find the exact shock wave solution of IDG, one needs to find the form of wave profile function. For this purpose, let us consider the massless point particle travelling in the positive $z$ direction with momentum $p^\mu=\lvert p\rvert(\delta^\mu_t+\delta^\mu_z)$. The associated null source creating the shock-wave geometry can be described as $T_{uu}=\lvert p\rvert \delta(x_\perp)\delta(u)$.
For the shock-wave ansatz (\ref{ansatz}),  the only non-vanishing components of the Ricci tensor is
\begin{equation}
 R_{uu}=G_{uu}=-\frac{\delta(u)}{2}\frac{\partial^2}{\partial _\perp^2}g(x_\perp).
\end{equation}
On the other hand, the energy momentum tensor in Kerr-Schild form can be written as $T_{\mu\nu}=\lvert p\rvert \delta(x_\perp)\delta(u)\lambda_\mu\lambda_\nu$. Therefore, the null source coupled IDG field equations (\ref{difeqppfinal}) reduce to the much simpler form
\begin{equation}
\bigg[1+\alpha_c \bigg(\partial_\perp^{2}{\cal F}_{2}(\partial_\perp^{2})+2\partial_\perp^{2}{\cal F}_{3}(\partial_\perp^{2})\bigg)\bigg]\partial_\perp^{2}g(x_\perp)=-16\pi G\lvert p\rvert \delta(x_\perp).
\label{shockwaveeq}
\end{equation}
For the simplest choice of the form factors as in (\ref{formfactor}), Eq.(\ref{shockwaveeq}) becomes a modified Poisson type equation\footnote{Note that the equation \ref{shapiroeqn1} is also studied in \cite{Frolov:2015usa} for a head-on collision of ultra-relativistic particles at the linearized level. Here, we have shown that the non-linear field equations of IDG for the shock wave ansatz reduce to this linear form.}
\
\begin{equation}
\begin{aligned}
e^{-\frac{\partial_{\perp}^2 }{M_s^2}}\partial_{\perp}^2 g(x_{\perp})=-16\pi G\lvert p\rvert\delta(x_{\perp}).
\label{shapiroeqn1}
\end{aligned}
\end{equation}
After using Fourier transform and evaluating related integrals, the axial symmetric solution can be obtained as
\begin{equation}
g(r)=-8G\lvert p\rvert\bigg(\mbox{ln}(\frac{r}{r_0})-\frac{1}{2}\mbox{Ei}(-\frac{r^2M_s^2}{4})\bigg),
\end{equation}
where $r$ is the distance to the origin defined as $r=\sqrt{x_{\perp}^2}$ and $r_0$ is integral constant. Here, $\mbox{Ei}$ is the exponential integral function\footnote{Exponential integral function for negative arguments defined by the integral \cite{Gradshteyn,Abramowitz}
\begin{equation}
\mbox{Ei}(r)=-\int_{-r}^{\infty} \frac{e^{-t}}{t}dt,
\end{equation}
and its derivative is $\mbox{Ei}'(r)=\frac{d}{dr}\mbox{Ei}(r)=\frac{e^r}{r}$.}.
Note that in the $M_s\to \infty$ limit, the profile function becomes \cite{Bonnor:1969rb,shock, Dray}
\begin{equation}
g(r)=-8 G\lvert p\rvert \mbox{ln}(\frac{r}{r_0}),
\end{equation}
which is the Einstein's gravity result as expected. Thus, the exact gravitational shock wave solution metric for IDG is
\begin{equation}
ds^2=-du dv-4 G\lvert p\rvert\delta(u)\bigg(\mbox{ln}(\frac{r^2}{r^2_0})-\mbox{Ei}(-\frac{r^2M_s^2}{4})\bigg) du^2+dx_\perp^2.
\label{ansatzIDG}
\end{equation}
Note that there is a distributional term in the null coordinate $u$, but this discontinuity can be removed by redefining new coordinates \cite{Dray}. On the other hand, for small distances (in the UV regime of non-locality), since expanding the exponential integral function into Puiseux series around $r=0$ gives \cite{Gradshteyn,Harris} 
 \begin{equation}
\mbox{Ei}(r)= \gamma+\mbox{ln} \lvert r \rvert+r+{\cal{O}}(r^2),
 \end{equation}
where $\gamma$ is Euler-Mascheroni constant. In the non-local regime $M_s r\ll 2$, the profile function is non-singular and reduces to
\begin{equation}
\lim_{M_s r\to 0}g(r)= g_{0}=4 G\gamma \lvert p\rvert ,
\label{non-singularshock}
\end{equation}
which is a constant. Here, for the sake of simplicity we set $r_0=\frac{2}{M_s}$. It is important to note that this choice does not affect the result in (\ref{non-singularshock}) to be constant. Interestingly, gravitational shock wave solution of IDG is non-singular in the UV regime of non-locality $M_s r\ll2$ while the result of pure GR diverges. Even though shock wave is produced by null matter source which contains Dirac delta function type singularity in the radial direction, the solution is non-singular at the origin due to the improved behaviour of the propagator in the UV scale. 

In fact, the discussion given above is not enough to conclude that the singularity disappears. One must also analyse whether curvature tensor diverges at the origin or not. Even if some modified gravity theories which contain four derivatives or less such as quadratic gravity have non-singular profile function  \footnote{For regularity properties of higher derivative gravity theories which contain at least six derivatives, see \cite{Frolov:2015bia,Giacchini:2018gxp}.}, some component of Riemann tensor diverges logarithmically \cite{Lousto:1996ep,Campanelli:1995ex}. Now, let us show that curvature tensors and invariants are non-singular at the position of the particle for the non-singular metric (\ref{ansatzIDG}) in the ghost-free IDG. One can demonstrate that the only non-zero components of the Riemann tensor are
\begin{equation}
\begin{aligned}
&R^{v}\,_{rur}=8G\lvert p \rvert\delta(u)\bigg(\frac{(1-e^{-\frac{r^2}{4M_s^2}})}{r^2}-\frac{e^{-\frac{r^2}{4M_s^2}}}{2M_s^2}\bigg),\hskip .5 cm 
R^{v}\,_{\phi u\phi}=8G\lvert p \rvert\delta(u)(1-e^{-\frac{r^2}{4M_s^2}}),\\&
R^{\phi}\,_{u u\phi}=4G\lvert p \rvert\delta(u)\frac{(-1+e^{-\frac{r^2}{4M_s^2}})}{r^2},
\hskip .5 cm R^{r}\,_{uur}=4G\lvert p \rvert\delta(u)\bigg(\frac{(1-e^{-\frac{r^2}{4M_s^2}})}{r^2}-\frac{e^{-\frac{r^2}{4M_s^2}}}{2M_s^2}\bigg),
\end{aligned}
\end{equation} 
wherein  the components for the $M_s r\to 0$ limit behave as
\begin{equation}
R^{v}\,_{rur}\sim-\frac{2G\lvert p \rvert\delta(u)}{M_s^2},\hskip .5 cm 
R^{v}\,_{\phi u\phi}\sim 0,\hskip .5 cm 
R^{\phi}\,_{u u\phi}\sim-\frac{G\lvert p \rvert\delta(u)}{M_s^2}
\hskip .5 cm R^{r}\,_{uur}\sim-\frac{G\lvert p \rvert\delta(u)}{M_s^2},
\end{equation} 
which are finite at the origin. So, all the non-zero components of Riemann tensor are non-singular in the UV regime of non-locality $M_s r\ll2$. On the other hand, the only non-vanishing component of the Ricci tensor is
\begin{equation}
R_{uu}=2G\lvert p \rvert\delta(u)\frac{e^{-\frac{r^2}{4M_s^2}}}{M_s^2},
\end{equation}
which approaches to a constant in the non-local region. Finally, the scalar curvature vanishes, all components of Weyl tensor are zero  ($C_{\rho\sigma\mu\nu}\sim 0$) in the in $M_s r\to 0$ limit and all the curvature invariants squared are given by
\begin{equation}
R^2=0, \hskip .5 cm R_{\mu\nu}R^{\mu\nu}=0,\hskip .5 cm  {\cal{K}}=R_{\mu\nu\rho\sigma}R^{\mu\nu\rho\sigma}=0,\hskip .5 cm C_{\mu\nu\rho\sigma}C^{\mu\nu\rho\sigma}=0,
\label{CIS1}
\end{equation}
where ${\cal{K}}$ is the Kretschmann scalar. In fact, the results given in (\ref{CIS1}) are direct consequence of the fact that all the curvature scalars vanish for the pp-wave metric \cite{Peres,Pravda:2002us}. With this discussion, we have shown that the gravitational shock wave solution of IDG is non-singular at the origin. It is also important to note that to investigate the non-singular nature, one usually chooses a geodesic and construct a frame parallelly transported along the geodesic completeness \cite{Ellis,Coley:2009uf}. For this purpose, say $e^\mu_{(a)}$ are such parallelly transported frames, then one needs to compute $R_{abcd}=e^\mu_{(a)}e^\nu_{(b)}e^\rho_{(c)}e^\sigma_{(d)}R_{\mu\nu\rho\sigma}$ and show the finiteness of $R_{abcd}$ \footnote{We would like to thank the referee to bring this point to our attention.}. But, since its beyond scope of the core of the current study, we will not do this here.

\section{Conclusions}
In this work, we studied exact pp-wave metrics of the ghost and singularity-free IDG and showed that the exact $pp$-wave solutions of the source-free IDG theory are also solutions of Einstein's general relativity. The pp-wave metrics also solve linearized field equations of the IDG. That is, the field equations of non-linear theory coincide with the linearized field equations for the pp-wave metrics. Undoubtedly, finding exact solution is not easy task since the field equations of the theory are highly non-trivial and non-linear. But, writing the metric in the Kerr-Schild form leads to a remarkable simplification on the field equations. 

We have also concentrated on the special class of axisymmetric pp-waves. Here, we studied the non-perturbative solution of the theory in the presence of the null-source and found the exact non-singular gravitational shock wave solution of the theory. We have shown that unlike the case in Einstein's gravity, although gravitational shock wave solution are created by a source having Dirac delta type singularity, the solution and curvature tensors are regular in the non-local regime due to gravitational non-local interactions. Even though, some non-singular solutions of the IDG at the linearized level are known \cite{Mazumdar1,Mazumdar2,Mazumdar3}, we find a non-singular gravitational shock wave solution for the theory at the non-linear level.

Although, we considered the exact solutions in the ghost-free IDG with a zero cosmological constant, this work can be extended to the case of non-zero cosmological constant as was done for quadratic gravity \cite{Gursespp}. For example, AdS plane waves are potential exact solutions of the theory.  On the other hand, studying Kerr-Schild class of metrics in non-local gravity models \cite{Barvinsky:1985an,Deser:2007jk,Conroy:2014eja} which are the infrared modification of GR, where the form factors are non-analytic function's of d'Alembert operator, would also be interesting.
\section{\label{ackno} Acknowledgements}
We would like to thank B. Tekin and S. Dengiz for useful discussions, suggestions and critical readings of the manuscript. We would also like to thank T. de Paula Netto and I. Kol\'a\v{r}  for suggestions and comments.


\begin{thebibliography}{0}


  \bibitem{Biswas2} 
  T.~Biswas, A.~Mazumdar and W.~Siegel,
  ``Bouncing universes in string-inspired gravity,''
  JCAP {\bf 0603}, 009 (2006).
  
  \bibitem{Biswas1} 
  T.~Biswas, E.~Gerwick, T.~Koivisto and A.~Mazumdar,
  ``Towards singularity and ghost free theories of gravity,''
  Phys.\ Rev.\ Lett.\  {\bf 108}, 031101 (2012).
  
   \bibitem{Modesto1} 
  L.~Modesto,
  ``Super-renormalizable Quantum Gravity,''
  Phys.\ Rev.\ D {\bf 86}, 044005 (2012).
  
  \bibitem{Biswas:2013kla} 
  T.~Biswas, T.~Koivisto and A.~Mazumdar,
  ``Nonlocal theories of gravity: the flat space propagator,''
  arXiv:1302.0532 [gr-qc].
  
  
  \bibitem{Edholm:2016hbt} 
  J.~Edholm, A.~S.~Koshelev and A.~Mazumdar,
  ``Behavior of the Newtonian potential for ghost-free gravity and singularity-free gravity,''
  Phys.\ Rev.\ D {\bf 94}, no. 10, 104033 (2016). 
  
  \bibitem{Kilicarslan:2018yxd} 
  E.~Kilicarslan,
  ``Weak Field Limit of Infinite Derivative Gravity,''
  Phys.\ Rev.\ D {\bf 98}, no. 6, 064048 (2018).
  
  \bibitem{Talaganis} 
  S.~Talaganis, T.~Biswas and A.~Mazumdar,
  ``Towards understanding the ultraviolet behavior of quantum loops in infinite-derivative theories of gravity,''
  Class.\ Quant.\ Grav.\  {\bf 32}, no. 21, 215017 (2015).
  
   \bibitem{Talaganis:2016ovm} 
  S.~Talaganis and A.~Mazumdar,
  ``High-Energy Scatterings in Infinite-Derivative Field Theory and Ghost-Free Gravity,''
  Class.\ Quant.\ Grav.\  {\bf 33}, no. 14, 145005 (2016).
  
  \bibitem{Tomboulis} 
  E.~T.~Tomboulis,
  ``Superrenormalizable gauge and gravitational theories,''
  hep-th/9702146.
  
  \bibitem{Buoninfante:2018mre} 
  L.~Buoninfante, G.~Lambiase and A.~Mazumdar,
  ``Ghost-free infinite derivative quantum field theory,''
  arXiv:1805.03559 [hep-th].
  
  \bibitem{Calcagni:2014vxa} 
  G.~Calcagni and L.~Modesto,
  ``Nonlocal quantum gravity and M-theory,''
  Phys.\ Rev.\ D {\bf 91}, no. 12, 124059 (2015).
 
 
 \bibitem{Biswas4} 
  T.~Biswas, A.~Conroy, A.~S.~Koshelev and A.~Mazumdar,
  ``Generalized ghost-free quadratic curvature gravity,''
  Class.\ Quant.\ Grav.\  {\bf 31}, 015022 (2014).
  
  \bibitem{Modesto} 
  L.~Modesto, J.~W.~Moffat and P.~Nicolini,
  ``Black holes in an ultraviolet complete quantum gravity,''
  Phys.\ Lett.\ B {\bf 695}, 397 (2011).
  
  \bibitem{Biswas5} 
  T.~Biswas, T.~Koivisto and A.~Mazumdar,
  ``Towards a resolution of the cosmological singularity in non-local higher derivative theories of gravity,''
  JCAP {\bf 1011}, 008 (2010).
  
  \bibitem{Biswas6} 
  T.~Biswas, A.~S.~Koshelev, A.~Mazumdar and S.~Y.~Vernov,
  ``Stable bounce and inflation in non-local higher derivative cosmology,''
  JCAP {\bf 1208}, 024 (2012).
  
  \bibitem{Frolov:2015bta} 
  V.~P.~Frolov,
  ``Mass-gap for black hole formation in higher derivative and ghost free gravity,''
  Phys.\ Rev.\ Lett.\  {\bf 115}, no. 5, 051102 (2015).
  
  \bibitem{Koshelev:2017tvv} 
  A.~S.~Koshelev, K.~Sravan Kumar and A.~A.~Starobinsky,
  ``$R^2$ inflation to probe non-perturbative quantum gravity,''
  JHEP {\bf 1803}, 071 (2018).
  
  \bibitem{Koshelev:2018hpt} 
  A.~S.~Koshelev, J.~Marto and A.~Mazumdar,
  ``Schwarzschild $1/r$-singularity is not permissible in ghost free quadratic curvature infinite derivative gravity,''
  Phys.\ Rev.\ D {\bf 98}, no. 6, 064023 (2018).
  
  \bibitem{Boos:2018bxf} 
  J.~Boos, V.~P.~Frolov and A.~Zelnikov,
  ``Gravitational field of static p -branes in linearized ghost-free gravity,''
  Phys.\ Rev.\ D {\bf 97}, no. 8, 084021 (2018).
  
  \bibitem{Boos:2018bhd}
  J.~Boos,
  ``Gravitational Friedel oscillations in higher-derivative and infinite-derivative gravity?,''
  Int. J. Mod. Phys. D \textbf{27}, no.14, 1847022 (2018).
  
\bibitem{book} H.~Stephani, D.~Kramer, M.~MacCallum, C.~Hoenselaers,
and E.~Herlt, \emph{Exact Solutions of Einstein's Field Equations}
(Cambridge University Press, Cambridge, 2003).

\bibitem{Eddington} A.~S.~Eddington, \emph{The Mathematical Theory of Relativity} (Cambridge University Press, Cambridge, 1924).
 \bibitem{Buchdahl}   
   H.~A~Buchdahl,  ''On Eddington's higher order equations of the gravitational field,'' Proceedings of the Edinburgh Mathematical Society, {\bf 8(2)}, 89-94 (1948). 
   
   \bibitem{Buchdahl2}
  H.~A.~Buchdahl, ''On the gravitational field equations arising from the square of the gaussian curvature,'' Nuovo Cim, {\bf 23}, 141 (1962). 
   
  \bibitem{Gullu:2011sj} 
  I.~Gullu, M.~Gurses, T.~C.~Sisman and B.~Tekin,
  ``AdS Waves as Exact Solutions to Quadratic Gravity,''
  Phys.\ Rev.\ D {\bf 83}, 084015 (2011).
  
  \bibitem{Gurses:2012db} 
  M.~Gurses, T.~C.~Sisman and B.~Tekin,
  ``New Exact Solutions of Quadratic Curvature Gravity,''
  Phys.\ Rev.\ D {\bf 86}, 024009 (2012).
  
   \bibitem{Gurses:2015zia} 
  M.~Gurses, T.~C.~Sisman and B.~Tekin,
  ``Gravity Waves in Three Dimensions,''
  Phys.\ Rev.\ D {\bf 92}, no. 8, 084016 (2015). 
  
  \bibitem{Gursespp} 
  M.~Gurses, T.~C.~Sisman and B.~Tekin,
  ``AdS-plane wave and $pp$-wave solutions of generic gravity theories,''
  Phys.\ Rev.\ D {\bf 90}, no. 12, 124005 (2014).
 
  \bibitem{Gurses:2013jua} 
  M.~Gurses, T.~C.~Sisman, B.~Tekin and S.~Hervik,
  ``AdS-Wave Solutions of f(Riemann) Theories,''
  Phys.\ Rev.\ Lett.\  {\bf 111}, 101101 (2013).
  
  \bibitem{Gurses:2011fv} 
  M.~Gurses, T.~C.~Sisman and B.~Tekin,
  ``Some exact solutions of all $f(R_{/mu/nu}$ theories in three dimensions,''
  Phys.\ Rev.\ D {\bf 86}, 024001 (2012).
  
  \bibitem{Sotiriou:2008rp} 
  T.~P.~Sotiriou and V.~Faraoni,
  ``f(R) Theories Of Gravity,''
  Rev.\ Mod.\ Phys.\  {\bf 82}, 451 (2010).
  
  \bibitem{Li:2015bqa} 
  Y.~D.~Li, L.~Modesto and L.~Rachwal,
  ''Exact solutions and spacetime singularities in nonlocal gravity,''
  JHEP {\bf 1512}, 173 (2015).
  
  \bibitem{Mazumdar1} 
  A.~S.~Cornell, G.~Harmsen, G.~Lambiase and A.~Mazumdar,
  ``Rotating metric in nonsingular infinite derivative theories of gravity,''
  Phys.\ Rev.\ D {\bf 97}, no. 10, 104006 (2018).
  
  \bibitem{Mazumdar2} 
  L.~Buoninfante, G.~Harmsen, S.~Maheshwari and A.~Mazumdar,
  ``Nonsingular metric for an electrically charged point-source in ghost-free infinite derivative gravity,''
  Phys.\ Rev.\ D {\bf 98}, no. 8, 084009 (2018).
  
  \bibitem{Mazumdar3} 
  L.~Buoninfante, A.~S.~Koshelev, G.~Lambiase, J.~Marto and A.~Mazumdar,
  ``Conformally-flat, non-singular static metric in infinite derivative gravity,''
  JCAP {\bf 1806}, no. 06, 014 (2018).
  
  \bibitem{Buoninfante:2019swn} 
  L.~Buoninfante and A.~Mazumdar,
  ``Nonlocal star as a blackhole mimicker,''
  arXiv:1903.01542 [gr-qc].
  
  \bibitem{Gueven:1987ad} 
  R.~Gueven,
  ``Plane Waves in Effective Field Theories of Superstrings,''
  Phys.\ Lett.\ B {\bf 191}, 275 (1987).
  
  \bibitem{Amati} 
  D.~Amati and C.~Klimcik,
  ``Nonperturbative Computation of the Weyl Anomaly for a Class of Nontrivial Backgrounds,''
  Phys.\ Lett.\ B {\bf 219}, 443 (1989).
  
  \bibitem{Horowitz} 
  G.~Horowitz and A.~R.~Steif,
  ``Spacetime singularities in string theory,''
  Phys.\ Rev.\ Lett.\  {\bf 64} 260 (1990).
  
  \bibitem{Coley} 
  A.~A.~Coley, G.~W.~Gibbons, S.~Hervik and C.~N.~Pope,
  ``Metrics With Vanishing Quantum Corrections,''
  Class.\ Quant.\ Grav.\  {\bf 25}, 145017 (2008).
  
  \bibitem{Hervik}
  S.~Hervik, V.~Pravda and A.~Pravdova,
  ``Type III and N universal spacetimes,''
  Class.\ Quant.\ Grav.\  {\bf 31} (2014) 
  
  \bibitem{Blau} M.~Blau, ``Plane Waves and Penrose Limits'', Lecture
Notes for the ICTP School on Mathematics in String and Field Theory
(June 2-13 2003).

\bibitem{Classification} M.~Ortaggio, V.~Pravda and A.~Pravdova,
Class.\ Quantum Grav.\textbf{\ 30} 013001, (2013).

\bibitem{Edelstein:2016nml} 
  J.~D.~Edelstein, G.~Giribet, C.~Gomez, E.~Kilicarslan, M.~Leoni and B.~Tekin,
  ``Causality in 3D Massive Gravity Theories,''
  Phys.\ Rev.\ D {\bf 95}, no. 10, 104016 (2017).
  
  \bibitem{Peres} 
  A.~Peres,
  ``Some Gravitational Waves,''
  Phys.\ Rev.\ Lett.\  {\bf 3}, 571 (1959).
  
  \bibitem{Pravda:2002us} 
  V.~Pravda, A.~Pravdova, A.~Coley and R.~Milson,
  ``All space-times with vanishing curvature invariants,''
  Class.\ Quant.\ Grav.\  {\bf 19}, 6213 (2002).
  
   \bibitem{BarnabyKamran}
  N.~Barnaby and N.~Kamran, ``Dynamics with Infinitely Many Derivatives: Variable Coefficient Equations,'' JHEP \textbf{12}, 022 (2008).
  
  \bibitem{Kilicarslan2}
  S.~Dengiz, E.~Kilicarslan, I.~Kol\'a\v{r} and A.~Mazumdar,
  ``Impulsive waves in ghost free infinite derivative gravity in anti-de Sitter spacetime,''
  Phys. Rev. D \textbf{102}, no.4, 044016 (2020).
  
  \bibitem{book2} J.~B.~Griffiths and J.~Podolsk\'y, \emph{Exact
  	Space-Times in Einstein's General Relativity} (Cambridge University
  Press, Cambridge, 2009).
  
   \bibitem{deser_tekin} 
  S.~Deser and B.~Tekin,
 ``Energy in generic higher curvature gravity theories,''
  Phys.\ Rev.\ D {\bf 67}, 084009 (2003).
  
   \bibitem{book3} R. ~Penrose,\emph{In General relativity, Papers in Honour of J. L. Synge; O'Raifeartaigh, L.; Ed. Oxford:
  	Clarendon Press}, Oxford, 1972, pp. 101-115.
  
  \bibitem{Frolov:2015usa}
  V.~P.~Frolov and A.~Zelnikov,
  ``Head-on collision of ultrarelativistic particles in ghost-free theories of gravity,''
  Phys. Rev. D \textbf{93}, no.6, 064048 (2016).
  
  \bibitem{Gradshteyn}
  I.~S. Gradshteyn and I.~M. Ryzhik,
  \newblock {\emph{ Table of {I}ntegrals, {S}eries, and {P}roducts}}.
  \newblock Edited by A. Jeffrey and D. Zwillinger. Academic Press, New York, 7th
  edition, 2007.
  
  \bibitem{Abramowitz}
  Abramowitz M and Stegun I A: \emph{ Handbook of Mathematical
  	Functions} (New York: Dover Publications) (1972). 
  
  \bibitem{Dray} 
  T.~Dray and G.~'t Hooft,
  ''The gravitational shock wave of a massless particle,''
  Nucl.\ Phys.\ B {\bf 253}, 173 (1985).
  
  \bibitem{Bonnor:1969rb} 
  W.~B.~Bonnor,
  ``The gravitational field of light,''
 Commun.\ Math.\ Phys. {\bf 13}, 163 (1969). 
 
  \bibitem{shock} 
P.~C.~Aichelburg and R.~U.~Sexl,
``On the gravitational field of a massless particle,''
Gen.\ Rel.\ Grav.\ {\bf 2}, 303 (1971).


\bibitem{Harris}
F.~E.~Harris, "Tables of the exponential integral Ei(x)," Math.\ Comp., v. 11, 1957, pp. 9-16. MR 19,464.


\bibitem{Frolov:2015bia} 
  V.~P.~Frolov, A.~Zelnikov and T.~de Paula Netto,
  ``Spherical collapse of small masses in the ghost-free gravity,''
  JHEP {\bf 1506}, 107 (2015).
  
  \bibitem{Giacchini:2018gxp} 
  B.~L.~Giacchini and T.~de Paula Netto,
  ``Weak-field limit and regular solutions in polynomial higher-derivative gravities,''
  arXiv:1806.05664 [gr-qc].

  \bibitem{Campanelli:1995ex} 
  M.~Campanelli and C.~O.~Lousto,
  ``Exact gravitational shock wave solution of higher order theories,''
  Phys.\ Rev.\ D {\bf 54}, 3854 (1996).
  
  \bibitem{Lousto:1996ep} 
  C.~O.~Lousto and F.~D.~Mazzitelli,
  ``Exact selfconsistent gravitational shock wave in semiclassical gravity,''
  Phys.\ Rev.\ D {\bf 56}, 3471 (1997).
  
   \bibitem{Ellis} 
   C.~B.~Collins and G.~F.~R.~Ellis, Phys.\ Rep.\ {\bf 56}, 65 (1979).
  
  \bibitem{Coley:2009uf} 
  A.~A.~Coley, S.~Hervik, W.~C.~Lim and M.~A.~H.~MacCallum,
  ``Properties of kinematic singularities,''
  Class.\ Quant.\ Grav.\  {\bf 26}, 215008 (2009).
  
  
   \bibitem{Barvinsky:1985an} 
  A.~O.~Barvinsky and G.~A.~Vilkovisky,
  ``The Generalized Schwinger-Dewitt Technique in Gauge Theories and Quantum Gravity,''
  Phys.\ Rept.\  {\bf 119}, 1 (1985).
  
  \bibitem{Deser:2007jk} 
  S.~Deser and R.~P.~Woodard,
  ``Nonlocal Cosmology,''
  Phys.\ Rev.\ Lett.\  {\bf 99}, 111301 (2007).
  
  \bibitem{Conroy:2014eja} 
  A.~Conroy, T.~Koivisto, A.~Mazumdar and A.~Teimouri,
  ``Generalized quadratic curvature, non-local infrared modifications of gravity and Newtonian potentials,''
  Class.\ Quant.\ Grav.\  {\bf 32}, 015024 (2015).
  
  
  
\end{thebibliography}
\end{document}